\documentclass{PoS}
\newcommand{\p}{\partial}
\newcommand{\Z}{{\mathbb{Z}}}

\title{String effects in Yang-Mills theory}

\ShortTitle{String effects in Yang-Mills theory}

\author{\speaker{M. Pepe} \\
INFN, Sezione di Milano-Bicocca\\ 
Edificio U2, Piazza della Scienza 3, 20126 Milano, Italy. \\
        E-mail: \email{pepe@mib.infn.it}}

\abstract{We study some features of the confining string connecting a quark-anti-quark
  pair in Yang-Mills theory. Monte Carlo investigations of the flux tube between two
  static quarks in the fundamental representation show that its thickness increases with
  the separation of the sources. The collected numerical data in (2+1)-d $SU(2)$
  Yang-Mills theory are in very good quantitative agreement with the next-to-leading order
  formula derived from the systematic low-energy effective theory of the confining
  string. At zero temperature, we observe the predicted logarithmic broadening. At finite
  and low temperature, the flux tube thickness is expected to broaden linearly. We also
  verify that prediction, finding an excellent agreement with the analytic expression. No
  adjustable parameter has been used to fit the numerical data.  Then we investigate the
  confining strings connecting color sources in larger representations of the gauge
  group. Concerning stable strings -- the $k$-strings -- we study the L\"uscher term of
  the fundamental string and of the 2-string in (2+1)-d $SU(4)$ Yang-Mills theory. We
  find that, at large separation between the two static sources, the coefficient of the
  L\"uscher term of the 2-string appraoches the value $-\pi/24$. However, at intermediate
  distances, there are relevant deviations. This result may suggest that, for $SU(N)$ at
  large $N$, a different intermediate string regime could set in.  For unstable strings,
  in (2+1)-d $SU(2)$ Yang-Mills theory we investigate the decay between two different
  string states and the multiple decays of the confining string connecting color sources in
  large representations. The multiple decays result from the progressive partial screening
  of the color sources.}

\FullConference{The XXVIII International Symposium on Lattice Filed Theory\\
                 June 14-19,2010\\
                 Villasimius, Sardinia Italy}

\begin{document}

\section{Introduction}

The confinement of quarks is one of the most fundamental features of the strong
interactions. The origin of this important property gives rise to many open questions and
a satisfactory understanding of the non-perturbative dynamics leading to quark confinement
is still lacking. Non-perturbative phenomena are quite common in physics and they usually
emerge in strongly coupled systems. For example, in condensed matter physics, strongly
coupled electrons are responsible for quantum antiferromagnetism and superconductivity. It
is usually a very challenging problem to understand the dynamical mechanisms that are
responsible for non-perturbative effects also because collective modes play a crucial
role.

However, one may disentangle the mechanism that gives rise to some non-perturbative effect
from the consequences of the effect itself. In fact, one may assume that some not yet
understood mechanism determines a given ground state for the system and then one may
investigate what consequences follow. In this framework, the typical energy scale of the
non-perturbative phenomenon represents an upper bound for the low-energy effects that one
wants to study. In general, the symmetries of the ground state --- internal or related to
space-time --- are a subset of the symmetries of the Lagrangian. Hence the Goldstone
bosons resulting from the spontaneous symmetry breaking are the natural degrees of freedom
to take into account in the low-energy regime of the theory. Then one constructs the most
general low-energy effective Lagrangian respecting the internal and space-time symmetries of the
system. Some parameters will enter the low-energy Lagrangian: their explicit value
can be derived only from the underlying  quantum theory and they represent the high-energy
non-perturbative input for the low-energy dynamics.

What we have just described is the approach of effective field theories. It turns out to
be a very powerful tool in investigating strongly coupled systems and provides a way to
perform calculations in physical systems that would otherwise be unmanageable in an
analytic manner. For example, chiral Lagrangians describe pion physics and pion-nucleon
systems in a very accurate way. Similarly, effective field theories for magnons provide an
excellent description of the low-energy dynamics of quantum antiferromagnets.

In the early '80s, L\"uscher, Symanzik, and Weisz \cite{Lue80} proposed a low-energy
effective string description for the $d$-dimensional Yang-Mills theory. In the confined
phase, a color flux tube connects the static quark-anti-quark pair.  During its
time-evolution, it sweeps out a two-dimensional world-sheet that breaks the translation
invariance of the Yang-Mills theory. The $(d-2)$ Goldstone bosons resulting from that
breaking are the degrees of freedom of the low-energy effective description of the
string. The leading term of the effective action describes the dynamics of a thin string
in the free-field approximation. Two non-trivial results follow. First, the leading
correction to the linear static potential \cite{Lue81} is given by
\begin{equation}
V(r) = \sigma r -\frac{\pi (d-2)}{24\, r} +{\cal {O}} (1/r^3),
\end{equation}
where $r$ is the distance between the two static sources and $\sigma$ is the string
tension. The universal L\"uscher term is proportional to the number $(d-2)$ of transverse
directions in which the string can vibrate. Many numerical simulations have successfully
shown the validity of that prediction
\cite{has,Cas96a,Jug02,Lue02,Cas04,Cas06,Har06,Bri08,Ath08,Bra09}.  Second, the
cross-sectional area swept out by the flux tube thickness is not fixed but increases with
the distance between the sources \cite{Lue81a}. At leading-order, the width of the color flux
tube at its midpoint $r/2$ is
\begin{equation}
w^2_{lo}(r/2)= \frac{(d-2)}{2\pi\sigma} \log (r/r_0),
\end{equation}
where $r_0$ is a length scale that enters the effective string theory as a low-energy
parameter. The above expression for the width is again universal. It is valid for strings
in the Yang-Mills theory as well as for interfaces separating different phases in condensed
matter. 

The low-energy dynamics discussed above cannot be straightforwardly extended to color flux
tubes connecting sources transforming under larger representations of the gauge group. The
flux tube itself may then be unstable and decay as the two color sources are pulled
apart. The stable flux tubes --- known as $k$-strings --- may have a non-trivial internal
structure and it is interesting to investigate whether a low-energy effective string
description can be constructed. In fact, considering $SU(N)$ Yang-Mills theory at
$N\rightarrow \infty$ \cite{Shi05}, $k$-strings are expected to consist of $k$
non-interacting fundamental strings. Assuming this picture to be correct, then we have $k$
copies of the effective string theory of the flux tube connecting two fundamental
charges. However, at large but finite values of $N$, the $k$ fundamental strings are
weakly interacting. Then, at very low energy scales, one should switch to the effective
string theory of a single string. Furthermore, it may be inappropriate to consider
$k$-strings as bound states of $k$ fundamental strings. In this case, a single vibrating
string is the correct low-energy description. Some numerical investigations have been
carried out in order to shed some light on the low-energy regime of the flux tubes
connecting color charges transforming under larger representations of the gauge group
\cite{Bri08,Ath08,Lip08,Wel10}.

The paper is organized as follows. In Section 2 we discuss the width of the color
flux tube as a function of the distance between the static sources. Numerical results are
compared with the analytic predictions obtained in the low-energy effective string
theory. Then, in Section 3, we consider color charges transforming under larger
representations of the gauge group. Finally, Section 4 contains the conclusions.

\section{The width of the color flux tube}

The low-energy effective string theory of the flux tube connecting two fundamental color
charges predicts that the width increases with the distance between the sources. Let
$h(x,t)$ be the scalar field that characterizes the displacement of the effective string
out of the plane of the two heavy source worldlines. In 2+1 dimensions, the effective
string action at next-to-leading order is given by \cite{Lue04,aha09,aha10A,aha10B}
\begin{equation}
\label{action3d}
S[h] = \frac{\sigma}{2} \int_0^\beta dt \int_0^r dx \ \left[\p_\mu h \p_\mu h -
\frac{1}{4} \left(\p_\mu h \p_\mu h\right)^2 \right].
\end{equation}
Here the variables $x\in[0,r]$ and $t\in[0,\beta]$ parametrize the 2-d base-space.  Since
the string ends at the static quark charges, the field $h(x,t)$ obeys the Dirichlet
boundary conditions $h(0,t)=h(r,t)=0$. Hence boundary terms can be present in the
effective string theory; however, due to open-closed string duality, it turns out that they
eventually contribute only at sub-leading orders \cite{Lue04,aha10A,aha10B}. At
next-to-leading order, the width of the string at the midpoint of the separation between
the two static sources is given by \cite{Gli10b}
\begin{equation}\label{w2C}
w^2(r/2)=\left( 1 +\frac{4\pi f(\tau)}{\sigma r^2}\right) w^2_{lo}
(r/2)-\frac{f(\tau)+g(\tau)}{\sigma^2r^2},
\end{equation}
where
\begin{equation}
f(\tau)=\frac{E_2(\tau)-4E_2(2\tau)}{48}\, , \quad
g(\tau)=i \pi \tau \left(\frac{E_2(\tau)}{12}- q\frac {d}{dq} \right) 
\left( f(\tau)+\frac{E_2(\tau)}{16} \right)+\frac{E_2(\tau)}{96}.
\end{equation}
The leading-order expression for the width has the following expression
\begin{equation}
w^2_{lo}(r/2) = \frac{1}{2 \pi \sigma} \log\left(\frac{r}{r_0}\right) + 
\frac{1}{\pi \sigma} \log\left(\frac{\eta(2 \tau)}{\eta(\tau)^2}\right).
\end{equation}
The parameter $q=\mbox{e}^{2\pi i \tau}$ depends on the ratio $\tau = i\beta/(2r)$ and 
\begin{equation}
E_2(\tau) = 1 - 24 \sum_{n=1}^\infty \frac{n q^n}{1-q^n}~,
\qquad
\eta(\tau) = q^{\frac1{24}} \prod_{n=1}^\infty(1-q^n)~,
\label{e2}
\end{equation}
are the first Eisenstein series $E_2$ and the Dedekind $\eta$-function, respectively.
Those functions obey the following transformations under the inversion 
$\tau \to - \tau^{-1}$
\begin{equation}
\eta(\tau) = \frac{1}{\sqrt{-i\tau}} \eta(- \tau^{-1})~, \quad
E_2(\tau) = \frac{1}{\tau^2} \, E_2(- \tau^{-1}) - \frac{6}{i \pi \tau}~.
\end{equation}
Using these relations, the expression of eq.(\ref{w2C}) in the zero-temperature limit,
$\beta \gg r $, is readily converted into an expression in the limit $r \gg \beta$. In
this case, the temperature is still small, but the string length $r$ is much larger than
the inverse temperature $\beta$. The leading-order string width then takes the form
\cite{All08,Cas10}
\begin{equation}
w_{lo}^2(r/2) = \frac{1}{2 \pi \sigma} \log\frac{\beta}{4 r_0} +
\frac{1}{4\beta \sigma}\, r +{\cal{O}}(\mbox{e}^{-2\pi r/\beta}).
\label{linear}
\end{equation}
This result shows that, at non-zero temperature, the flux tube width increases linearly
with the distance between the external static quarks. The finite temperature linear string
broadening has been observed for fluctuating interfaces in the 3-d Ising model
\cite{All08,Cas10}; the $(3+1)$-d $SU(3)$ Yang-Mills theory has been investigated in
\cite{Bak10}.

The lattice formulation of the Yang-Mills theory is a powerful tool that allows us to
study the string dynamics from first principles. However, the numerical investigation of
the width of the color flux tube is a very challenging problem. Hence, it is convenient to
first consider the least problematical case of (2+1)-d $SU(2)$ Yang-Mills theory on a cubic
lattice of size $L\times L\times \beta$. The temporal extent $\beta$ determines the
inverse temperature. The partition function is given by

\begin{equation}
Z = \int {\cal D}U \ \exp(- S[U]) = \prod_{x,\mu} \int_{SU(2)} dU_{x,\mu} \
\exp(- S[U]),
\end{equation}
where $dU_{x,\mu}$ is the local gauge invariant Haar measure for a parallel 
transporter variable $U_{x,\mu} \in SU(2)$ located on the link $(x,\mu)$. We use
the standard Wilson plaquette action 
\begin{equation}
S[U] = - \frac{2}{g^2} \sum_{x,\mu,\nu} 
\mbox{Tr} [U_{x,\mu} U_{x+\hat\mu,\nu} U_{x+\hat\nu,\mu}^\dagger U_{x,\nu}^\dagger],
\end{equation}
where $g$ is the bare gauge coupling and $\hat\mu$ is a unit-vector pointing in the
$\mu$-direction. All dimensionful quantities are expressed in units of the lattice spacing
which is set to 1.

The Polyakov loop $\Phi_x$ represents the insertion of an external static quark at the spatial
position $x$ into the thermal bath of dynamical gluons
\begin{equation}
\Phi_x = \mbox{Tr} \left[\prod_{t=1}^\beta U_{x+t\hat 2,2}\right].
\end{equation}
The direction 2 is chosen to represent the Euclidean time direction. Inserting a pair of
static quarks separated by a distance $r$ along the 1-direction, the confining string sweeps
out a two-dimensional world-sheet that fluctuates in the transverse 3-direction. The
static potential $V(r)$ is obtained from the Polyakov loop 2-point function
\begin{equation}
\langle \Phi_0 \Phi_r \rangle = \frac{1}{Z} \int {\cal D}U \ \Phi_0 \Phi_r \ \exp(- S[U]) 
\sim \exp(-\beta V(r)),
\end{equation}
in the zero-temperature limit $\beta \rightarrow \infty$. The force between the two static
quarks is derived by taking the discrete derivative of the static potential
\begin{equation}
F(\bar r)= V(r+1)-V(r) = -\frac{1}{\beta}\, \log 
\frac{\langle \Phi_0 \Phi_{r+1} \rangle}{\langle \Phi_0 \Phi_r \rangle},
\end{equation}
where, writing explicitly the lattice spacing $a$, 
$\bar r = r + \frac{a}{2} + {\cal{O}}(a^2)$ is an improved definition of the point where
the force is computed \cite{Som93}. The width
of the vibrating string is extracted from the connected correlation function
\begin{equation}\label{obs}
C(x_3) = \frac{\langle \Phi_0 \Phi_r P_x \rangle}{\langle \Phi_0 \Phi_r \rangle}
- \langle P_x \rangle,
\end{equation}
of two Polyakov loops with a single plaquette 
\begin{equation}
P_x = \mbox{Tr} [U_{x,1} U_{x+\hat 1,2} U_{x+\hat 2,1}^\dagger U_{x,2}^\dagger],
\end{equation}
in the $1$-$2$ plane. The plaquette is located at the site $x = (r/2,t,x_3)$ and measures
the fluctuations of the color flux tube at the midpoint $r/2$ between the external
quark charges as a function of the transverse displacement $x_3$. Then the squared width of the
string is given by the second moment of the normalized correlation function
\begin{equation}
w^2(r/2) = \frac{\int dx_3 \ x_3^2\, C(x_3)}{\int dx_3 \ C(x_3)}.
\label{widthmeas}
\end{equation}

Eq.(\ref{obs}) shows why the measurement of the width is a demanding numerical problem. In
fact, it is obtained from the ratio of two exponentially small observables. The
L\"uscher-Weisz multi-level simulation technique \cite{Lue01} allows us to attain an
exponential reduction of the statistical fluctuations and makes the required numerical
study feasible. In the case of the 3-d Ising model --- equivalent to the 3-d $\Z (2)$
gauge theory --- the static sources can be considered as part of the Boltzmann weight and
numerical simulations can be carried out with high efficiency.

In the next two subsections accurate numerical evidence for the broadening of the color
flux tube with the distance are discussed both at zero \cite{Gli10a} and at finite
temperature \cite{Gli10c}.

\subsection{The logarithmic broadening at zero temperature}
\begin{figure}[htb]
\centering
\includegraphics[height=.42\textwidth,width=0.7\textwidth]{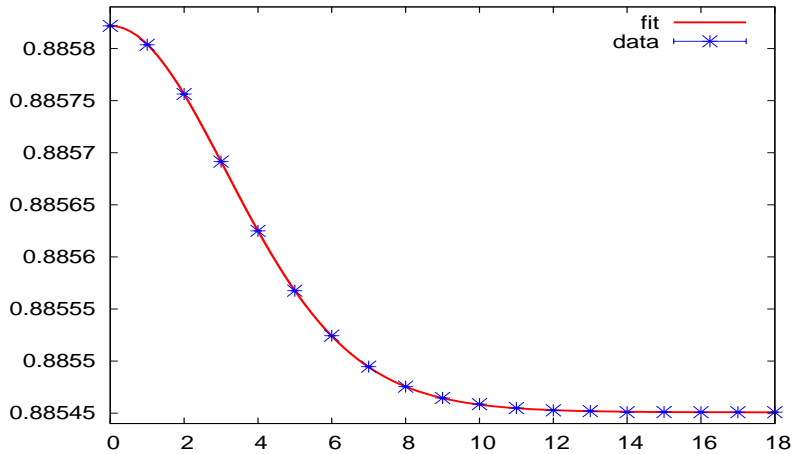}
\caption{The ratio $\frac{\langle \Phi_0 \Phi_r P_x \rangle}{\langle \Phi_0 \Phi_r \rangle}$
as a function of the transverse displacement  $x_3$ of the plaquette from the midpoint of
the interquark separation. The distance between the two static sources is $r=19$. The solid
curve is a fit using eq.(\protect\ref{fitbell}).}
\label{bell}\end{figure}

We consider the (2+1)-d $SU(2)$ Yang-Mills theory at coupling $4/g^2=9.0$ on a lattice
with $L=54$ and $\beta=48$. Figure \ref{bell} illustrates the correlation function of
eq.(\ref{obs}) at distance $r=19$ between the two static quarks. The data have the
expected bell-shape distribution. The high accuracy of the numerical data allows to detect
small deviations from a pure Gaussian and we fit using the ansatz
\begin{equation}\label{fitbell}
\frac{\langle \Phi_0 \Phi_r P_x \rangle}{\langle \Phi_0 \Phi_r \rangle} = A\, \exp(- x_3^2/s)\; 
\frac{1 + B\, \exp(- x_3^2/s)}{1 + D\, \exp(- x_3^2/s)} + K.
\end{equation}
The squared width is then determined from eq.(\ref{widthmeas}). The fit function of
eq.(\ref{fitbell}) has always provided a very good fit of the numerical data.  We have
checked that other choices of the fit function do not have a significant effect on the
value of the width as long as they give good fits.

\begin{figure}[htb]
\centering
\includegraphics[height=.44\textwidth,width=0.7\textwidth]{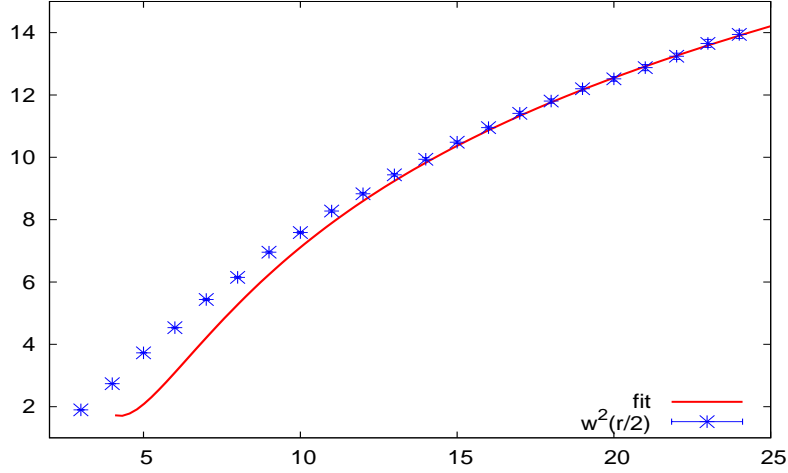}
\caption{The squared width $w^2(r/2)$ of the color flux tube at zero temperature evaluated
  at the midpoint between the static sources. The solid curve is a fit using the 2-loop
  expression of eq.(\protect\ref{w2C}) obtained from the effective string theory.}
\label{widthTzero}\end{figure}

Figure \ref{widthTzero} shows the dependence of the squared string width $w^2 (r/2)$
as a function of the distance $r$ between the static quarks. The solid curve is a fit of
the numerical data using eq.(\ref{w2C}). The data show that the flux tube thickness
indeed increases with the distance between the static quarks. Moreover, starting from
$r\sim 18$, the data are very well represented by the prediction of the effective string
theory. In particular, the leading order logarithmic broadening at zero temperature and the
predicted prefactor $1/(2\pi\sigma)$ turn out to be confirmed by the data.

After determining the value of the string tension $\sigma=0.025897(15)$ from the static
potential, there is only one free parameter in eq.(\ref{w2C}): from the fit, we obtain
$r_0=2.26(2)$.

\begin{figure}[htb]
\centering
\includegraphics[height=.44\textwidth,width=0.7\textwidth]{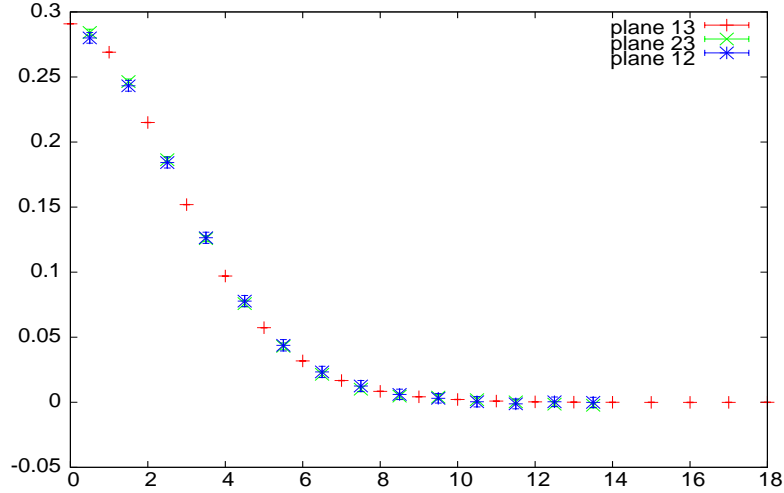}
\caption{Normalized probability distribution $C(x_3)/\int dx_3 C(x_3)$ using the three
  possible orientations for the plaquette in eq.(\protect\ref{obs}).}
\label{plaqorient}\end{figure}

In the numerical results discussed so far, the plaquette orientation is parallel to the plane
of the quark worldlines. However, one can also consider other orientations for the
plaquette in eq.(\ref{obs}). Figure \ref{plaqorient} shows the normalized probability
distribution for the confining string using the three different orientations for the
plaquette. The distance between the static sources is $r=12$.

\subsection{The linear broadening at finite temperature}
In this subsection we investigate the dependence of the width of the flux tube on
the distance between the static quarks at finite temperature. We again consider 
(2+1)-d $SU(2)$ Yang-Mills theory at $4/g^2=9.0$. The lattice size is now $90^2\times
16$ which corresponds to a temperature of about $0.38\; T_c$, where $T_c$ is the
temperature of the deconfinement phase transition.

It is important to note that the two low-energy constants, $\sigma$ and $r_0$, appearing
in eq.(\ref{w2C}) for the next-to-leading expression of the confining string width, have
been already fixed by a fit to the data obtained at zero temperature. Hence, there is no
longer an adjustable parameter to fit to the finite temperature numerical data and the
theoretical prediction is completely determined. This represents a very stringent test of
the validity of the low-energy effective string description of the confining string.

\begin{figure}[htb]
\centering
\includegraphics[height=0.44\textwidth,width=0.7\textwidth]{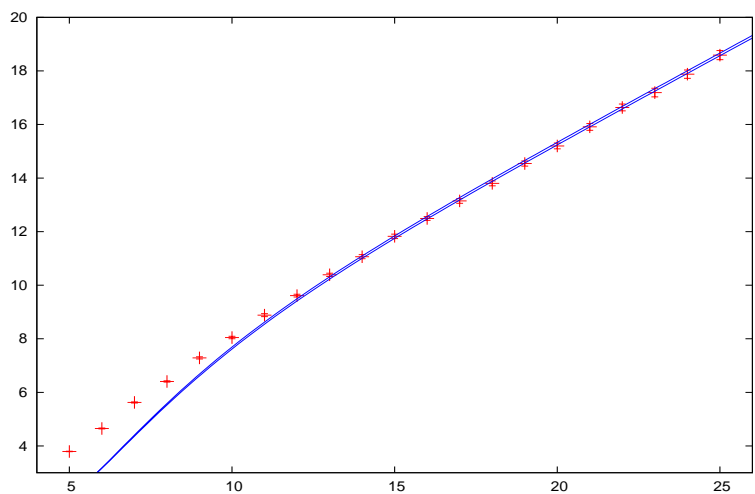}
\caption{The finite temperature squared width $w^2(r/2)$ of the color flux tube at the
  midpoint between the static sources. The solid curves are the 2-loop expression of
  eq.(\protect\ref{w2C}) obtained from the effective string theory using $r_0=2.26\pm 0.02$.}
\label{widthTfinite}\end{figure}

Figure \ref{widthTfinite} shows the dependence of the flux tube width on the separation
between the two static quarks as it results from the numerical simulations. The two solid
curves are the fully determined prediction eq.(\ref{w2C}) of the low-energy effective
string theory using the two choices $r_0=2.26\pm 0.02$. The numerical data show the
predicted linear broadening for sufficiently large separations. The agreement with the
analytic expression is excellent: this confirms in a quantitative manner the correctness of
the description of the low-energy dynamics of the flux tube as a fluctuating string.

\section{The confining string connecting color charges in larger representations}

In Yang-Mills theory the gluons are the only dynamical degrees of freedom. One can probe
their dynamics by inserting into the system external static color charges that
transform under a given representation of the gauge group. Since gluons transform under
the adjoint representation of the gauge group, $N$-ality is a relevant concept to
understand the behavior of color charges transforming under a generic representation of
the gauge group. Let us consider the gauge group $SU(N)$. Two representations ${\cal
  {R}}_1$ and ${\cal {R}}_2$ are in the same $N$-ality sector if the second representation
${\cal {R}}_2$ appears in the tensor product decomposition of ${\cal {R}}_1$ with a given
number of adjoint representations. This means that two color charges that transform under
the representations ${\cal {R}}_1$ and ${\cal {R}}_2$ can transmute into one another by
gluon exchange. For $SU(N)$ there are $N$ sectors: these $N$-ality sectors are not
independent but are pairwise related by charge conjugation symmetry (real and pseudoreal
representations are self-related).

Let us now insert into the system a quark-anti-quark pair transforming under a given
representation of the gauge group. The color flux tube that connects the two color sources
is, in general, unstable and it decays when the two charges are pulled apart to a
sufficiently large distance \cite{Pou95,Ste99,Phi99,Kra03,Pep09}. In every $N$-ality sector there is just a
single stable flux tube: it is the one with the minimal value of the string tension. These
stable strings are known as $k$-strings, where $k=0,\ldots, N-1$ labels the $N$-ality
sector.

The internal structure of the flux tube connecting color sources transforming under larger
representations of the gauge group is not known. Hence, it is not clear how to extend the
low-energy effective theory of the confining string connecting two fundamental
charges to those cases.

For $N\rightarrow \infty$, $k$-strings are expected to consist of $k$ non-interacting
fundamental strings. In that limit, the low-energy regime of $k$-strings is described by
$k$ replicas of the effective string connecting two fundamental color charges. For large,
but finite $N$, it is reasonable to expect that the $k$ fundamental strings are weakly
interacting. If $\Lambda$ is the energy scale of such an interaction, then one would
expect to observe two different low-energy regimes. In the energy range $[\Lambda, \sqrt
  \sigma]$ --- where $\sigma$ is the fundamental string tension --- the effective picture
of $k$ non-interacting fundamental strings may be a meaningful description. However, for
energies smaller than $\Lambda$, the bunch of $k$ fundamental strings merge and
effectively behave like a single flux tube. Then it is appropriate to consider the
effective theory of a single string with the string tension of the $k$-string as a
low-energy parameter. Moreover, when $N$ is not large, it may not be justified to view the
$k$-string as a composite object of fundamental strings and it is not clear what the
appropriate low-energy description of the confining flux tube is.

In the next two subsections, we report on numerical results of Monte Carlo simulations of
Yang-Mills theory with color sources in larger representations of the gauge group. In the
first subsection results for $k$-strings are discussed. In the second subsection the
decay of unstable strings is investigated.

\subsection{The effective string theory for $k$-strings}

In this subsection we present the results of numerical investigations on string effects
for $k$-strings. In $SU(2)$ and $SU(3)$ Yang-Mills theories, the confining string
connecting two fundamental color charges is the only stable string. Here we consider
$SU(4)$ Yang-Mills theory because it is the simplest pure gauge theory with a second
stable confining string. In fact the 2-string connecting two color charges in the $\{6\}$
representation of $SU(4)$ cannot break and it does not belong to the $N$-ality sector of
the fundamental $\{4\}$ representation. Since it is least problematical for numerical
simulations, we perform this study in (2+1) dimensions: we consider a $32^3$ lattice at
$8/g^2=21.$ In order to extract informations on the low-energy regime of the 2-string, we
measure the L\"uscher term. We expect to observe one of the following cases. If the
2-string consists of two non-interacting fundamental strings, then the coefficient of the
L\"uscher term should be twice the coefficient of the string connecting two
fundamental charges. If the two fundamental strings are strongly interacting or,
eventually, if the 2-string cannot even be interpreted as a bound state of fundamental
strings, then the L\"uscher term should be the same as the one of the fundamental
string. However, also other results may be obtained \cite{Giu07}. Numerical results of Teper
and collaborators \cite{Bri08,Ath08} support that the L\"uscher term of the 2-string is the same as
the one of the fundamental string. Here we perform an accurate comparison of the L\"uscher
term for the fundamental string and the 2-string. In particular we focus on the dependence
on the separation between the two static sources.

\begin{figure}[htb]
\centering
\includegraphics[height=.44\textwidth,width=0.7\textwidth]{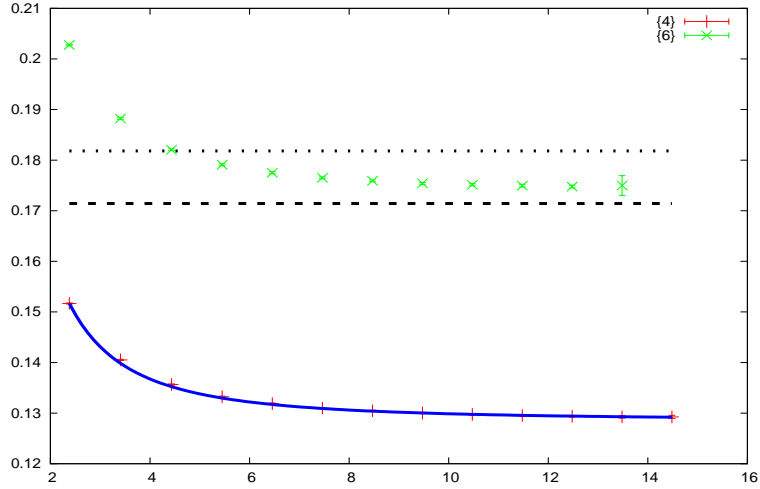}
\caption{The force exerted by the fundamental string and by the 2-string as a function of the
  separation between the color sources. The solid curve is a fit of the form 
$F(r) =  \sigma + \pi/(24 r^2)$. The dashed and dotted lines are the expected values for
  the string tension of the 2-string assuming the Casimir law and the sine law, respectively.}
\label{forceSU4}\end{figure}

Figure \ref{forceSU4} shows the force $F(r)$ as a function of the distance, both for the
fundamental string and the 2-string. The solid curve is a fit of the form $F(r) = \sigma +
\pi/(24 r^2)$ of the data for the fundamental string. We obtain the value $\sigma=0.12857(5) $
for the fundamental string tension. Then the dashed and the dotted lines correspond to the
predictions of the string tension of the 2-string assuming the Casimir law \cite{Amb84A,Amb84B} and the
sine law \cite{Han97}, respectively. The data for the force of the 2-string happen to be
in-between those two laws.

\begin{figure}[htb]
\centering
\includegraphics[height=.44\textwidth,width=0.7\textwidth]{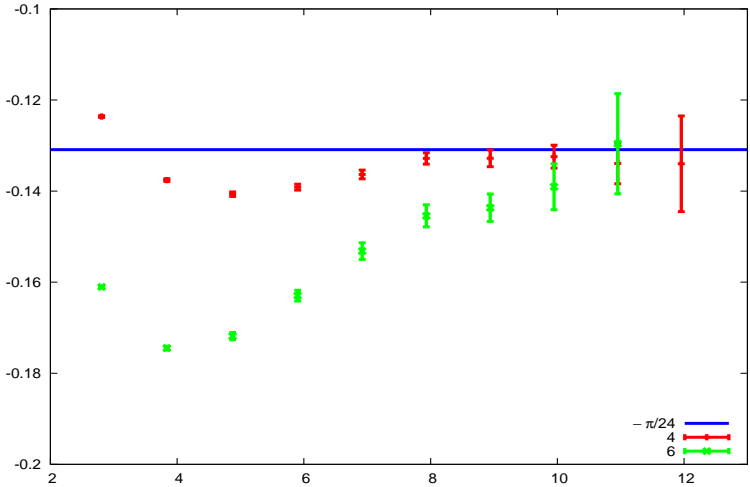}
\caption{The quantity $c(r)$ defined in eq.(\protect\ref{cr}) as a function of the separation of
  the color charges for the fundamental and the 2-string. The solid line is $-\pi/24$.}
\label{LuscherSU4}\end{figure}

Figure \ref{LuscherSU4} reports the value of $c(r)$ for the fundamental string and the
2-string
\begin{equation}
c(r)=\frac{1}{2} r^3 \; \frac{d^2 V(r)}{dr^2}.
\label{cr}
\end{equation}
For large separations between the color sources, the quantity $c(r)$ approaches
asymptotically the coefficient of the L\"uscher term.  The data for the 2-string
accurately confirm that, for sufficiently large distances, the 2-string behaves like a
single bosonic string: the quantity $c(r)$ indeed approaches the value $-\pi/24$ just like
the fundamental string. However, it turns out that for not very large distances, there
are important deviations from the asymptotic value. This might suggest that, as $N$
becomes large, the approach to $-\pi/24$ happens at larger and larger distances and a
different regime may eventually appear at intermediate separations between the static
sources.

\subsection{String decay}
In this subsection, we investigate the behavior of unstable strings as the distance
between the static color sources is increased. For simplicity, we consider the $SU(2)$
Yang-Mills theory in (2+1) dimensions. The only stable string is the one connecting two
color charges in the fundamental representation: all other confining strings either decay
into the fundamental one or break.  Adjoint color sources --- i.e. charges in the $\{3\}$
representation of $SU(2)$ --- belong to the $N$-ality sector of the trivial
representation. As we pull apart a pair of adjoint sources, the confining string that connects
the two static sources is being stretched. However, at sufficiently large distances, the string breaks and
the static potential flattens \cite{Pou95,Ste99,Phi99,Kra03}.

Color sources in the $\{4\}$ representation cannot be completely screened by $SU(2)$
gluons. Hence the confining string connecting two $\{4\}$ charges does not break
completely but decays into the fundamental string. The string connecting two charges in
the $\{5\}$ representation ultimately breaks. However, at intermediate distances,
dynamical gluons can partially screen the $\{5\}$ charges so that they behave as effective
color charges in the adjoint representation. Thus, the initial string can undergo multiple
decays with a progressive partial screening at every decay step \cite{Pep09}.
 
\begin{figure}[htb]
\centering
\includegraphics[height=.4\textwidth,width=0.5\textwidth]{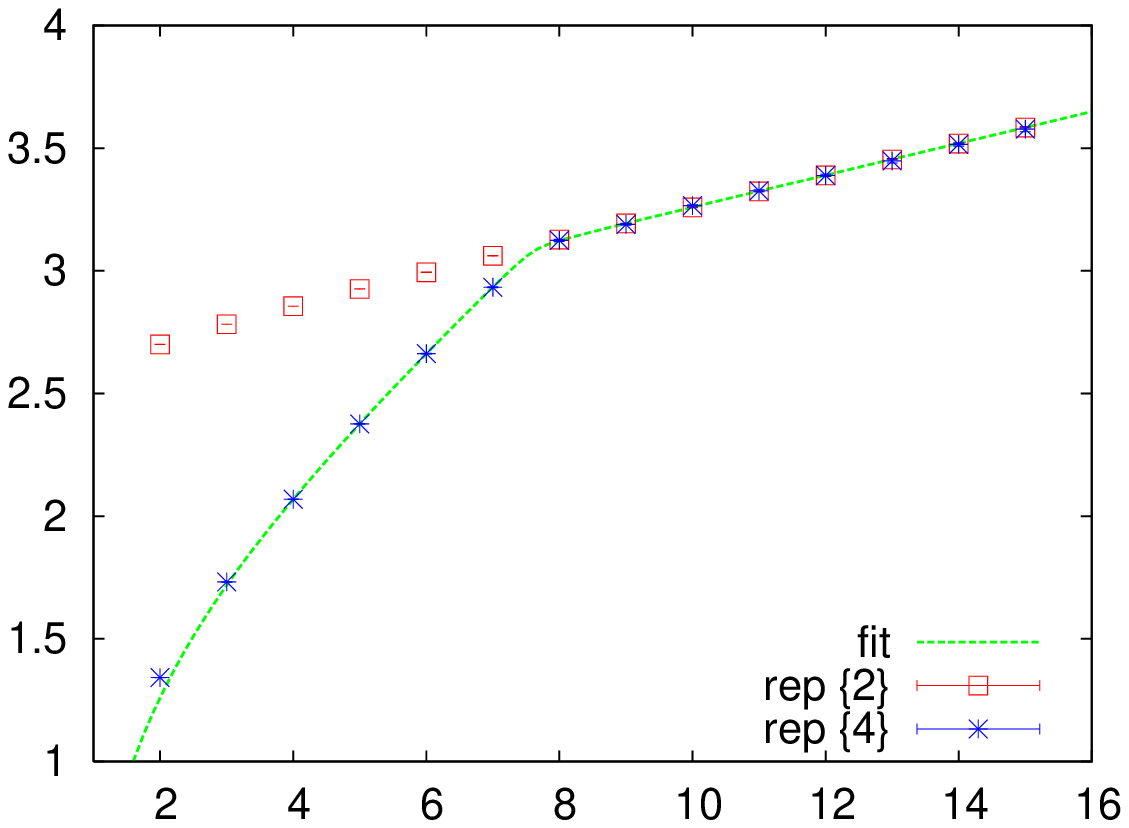}
\hskip-.1cm
\includegraphics[height=.4\textwidth,width=0.5\textwidth]{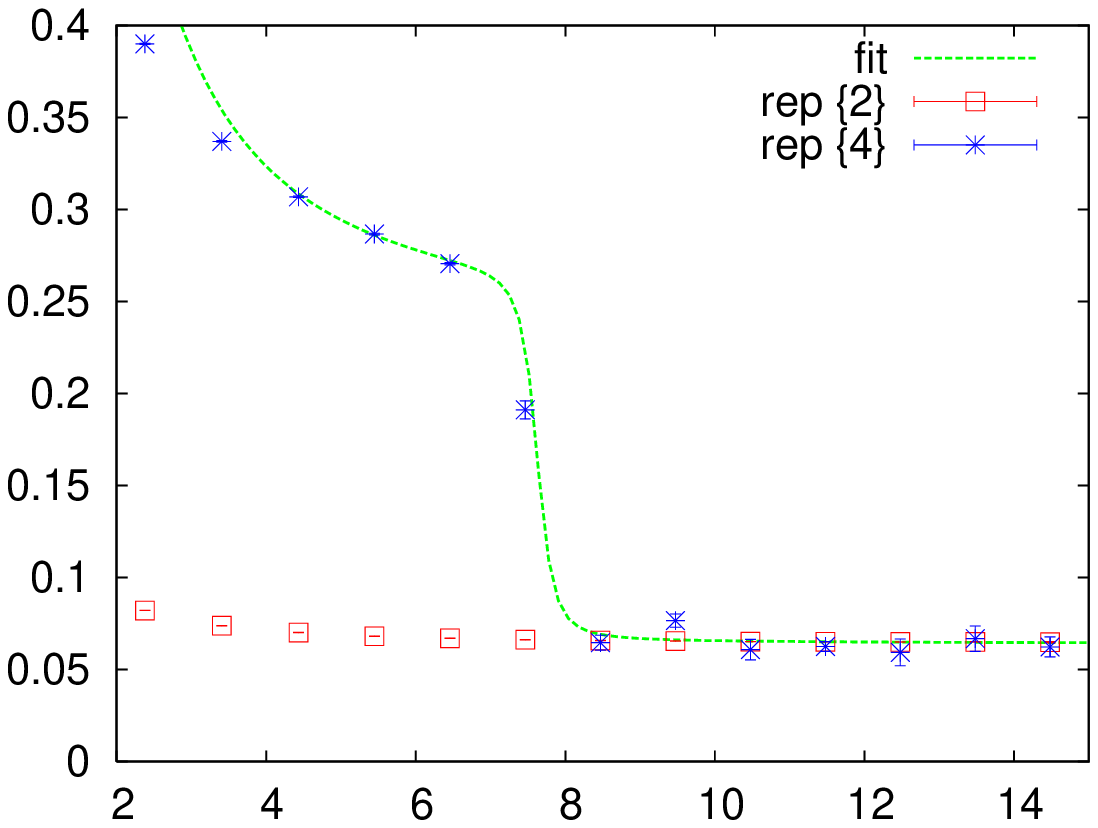}
\caption{The potential (left) and the force (right) for charges in the $\{2\}$
and in the $\{4\}$ representations.}
\label{decayhalfinteger}\end{figure}

We show here the data of numerical simulations of (2+1)-d $SU(2)$ Yang-Mills theory on a
$32^2\times 64$ lattice at coupling $4/g^2=6.0$. The left-panel of figure
\ref{decayhalfinteger} compares the static potentials of two $\{4\}$ charges and two
fundamental $\{2\}$ charges. At distances $r\sim 8$, the $\{4\}$ charges get screened:
they behave as two fundamental charges and the string decays. In fact, after the string
decay, the $\{4\}$ static potential is the same as that of two fundamental charges. In the
right-panel of figure \ref{decayhalfinteger} we display the results for the force. The
data give evidence that, after the decay of the $\{4\}$ string, the string tension takes the
value of the fundamental string tension.

\begin{figure}[htb]
\centering
\includegraphics[height=.4\textwidth,width=0.5\textwidth]{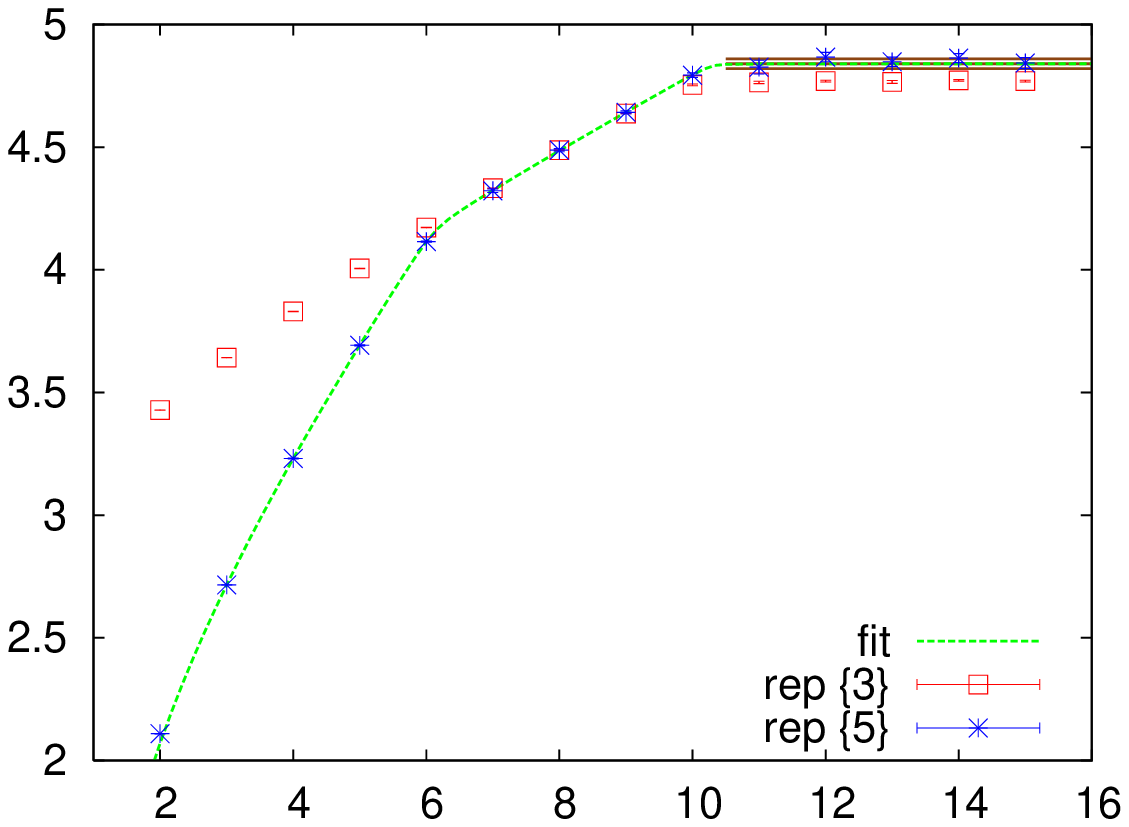}
\hskip-.1cm
\includegraphics[height=.4\textwidth,width=0.5\textwidth]{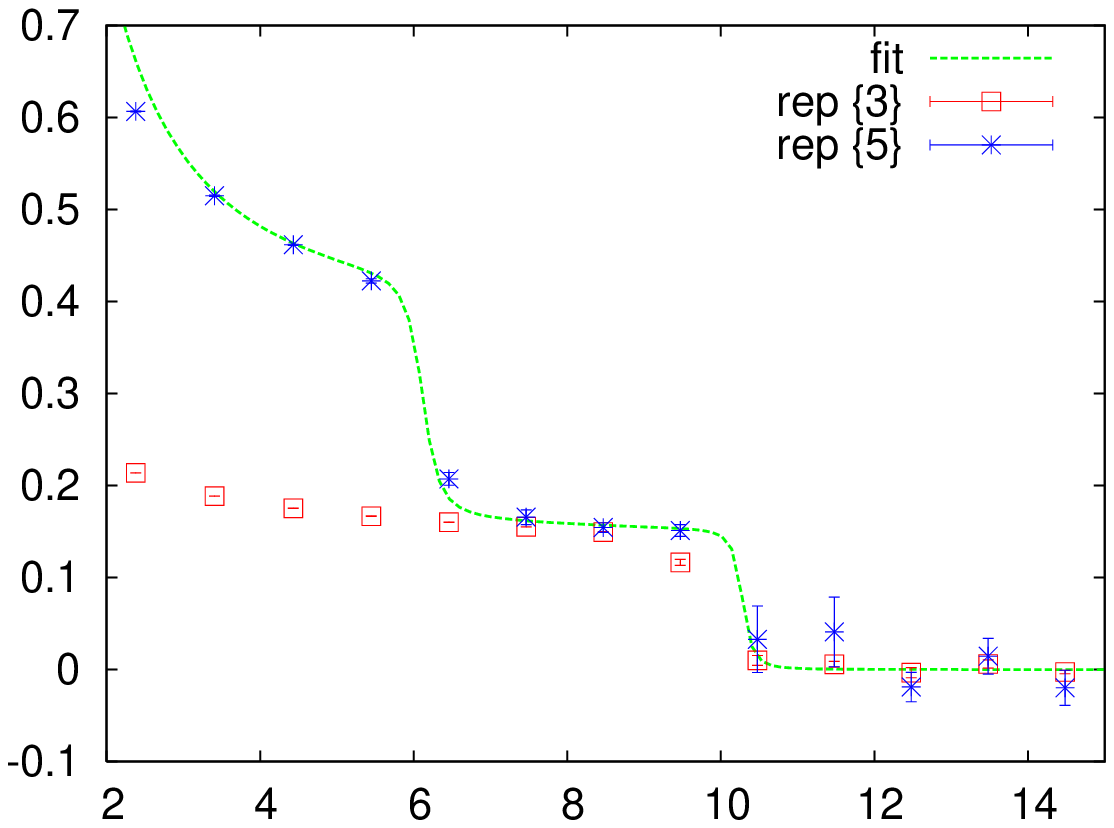}
\caption{The potential (left) and the force (right) for charges in the $\{3\}$
and in the $\{5\}$ representations. The band in the left-panel corresponds to twice the free energy of an
isolated quark in the $\{5\}$ representation.}
\label{decayinteger}\end{figure}

The left-panel of figure \ref{decayinteger} reports the results for the static potential
of $\{5\}$ and $\{3\}$ color charges. The data for the $\{5\}$ representation show that
indeed charge screening takes place in partial steps. The $\{5\}$ charge is first
partially screened to behave as an effective $\{3\}$ charge: this is what the first kink
in the static potential shows. As we can see in the right-panel of figure
\ref{decayinteger}, after the first decay of the $\{5\}$ string, the value of the force
agrees with the force between two $\{3\}$ charges. At the second decay the string breaks
completely and the string tension vanishes.

Interestingly, there is a small difference between the flattening levels of the $\{3\}$
and of the $\{5\}$ static potentials. This is most probably due to the different gluon
content of the doubly screened $\{5\}$ charge and of the singly screened $\{3\}$ charge.

It is interesting to extend the investigation of the string decay also to other groups
than $SU(N)$. For instance $Sp(N)$ groups \cite{Hol03A} have only two $N$-ality sectors
like $SU(2)$. The fundamental representation of $Sp(2)$ is $\{4\}$ and $\{10\}$ is the
adjoint representation.  Interestingly, $Sp(2)$ has a center-neutral representation with
smaller dimensionality than the adjoint: it is $\{5\}$, the fundamental representation of
$SO(5)$. Since the decomposition of the tensor product of $\{5\}$ and the adjoint
representation is given by
\begin{equation}
\{5\} \otimes \{ 10\} = \{5\} \otimes \{ 10\} \oplus \{ 35\},
\end{equation}
the $\{5\}$ charge needs two $Sp(2)$ gluons to be completely screened. Moreover it is
reasonable to expect that the string tension between two $\{5\}$ charges is smaller than
the one between two adjoint $\{10\}$ charges: thus, the $\{5\}$ string breaks in a single step with
the simultaneous dynamical production of 4 $Sp(2)$ gluons.

Another interesting example is related to the fundamental $\{7\}$ charge of $G(2)$. In $G(2)$
Yang-Mills theory \cite{Hol03B} there are no stable strings and all confining strings
ultimately break.  A fundamental $\{7\}$ charge is completely screened by 3 $G(2)$ gluons,
belonging to the representation $\{14\}$. However, the fundamental $\{7\}$ representation
is the smallest representation of the group $G(2)$ and the confining string connecting two
$\{7\}$ charges has the minimal string tension \cite{Lip08}. Thus, partial screening
of the fundamental $\{7\}$ charge cannot happen: the string breaks in just one step by
popping out from the vacuum 6 $G(2)$ gluons \cite{Wel10}.

\section{Conclusions}

In this paper we have presented recent results in the investigation of string effects in
Yang-Mills theory. We have discussed confining strings connecting fundamental charges
as well as color sources transforming under larger representations of the gauge group.

The color flux tube connecting two fundamental charges broadens as the separation of the
static sources is increased. We have performed numerical simulations in (2+1)-d $SU(2)$ Yang-Mills
theory both at zero and at finite temperature. At zero temperature we have shown that the
width of the confining string increases logarithmically, in very good agreement with
the analytic results obtained in the low-energy effective string theory for the color flux
tube. The 2-loop expression for the width obtained using the low-energy effective string
theory involves two low-energy constants: the string tension $\sigma$ and a distance scale
$r_0$. The string tension has been fixed using the static potential and, thus, the fit of
the numerical data for the width has only one free parameter.

The low-energy effective string theory also predicts that the flux tube broadens linearly
at finite temperature. The numerical simulations have confirmed this expectation to very
high accuracy. It is remarkable that, after having fixed the value of $\sigma$ and $r_0$
using the zero temperature results, the prediction obtained in the low-energy effective
string theory is completely determined.

Then the confining strings connecting color charges transforming under larger
representations of the gauge group have been investigated. For stable strings --- the
so-called $k$-strings --- we have investigated the L\"uscher term. We have performed
numerical simulations of (2+1)-d $SU(4)$ Yang-Mills theory and measured with high accuracy
the static potential for $\{4\}$ and $\{6\}$ color charges. The coefficient of the
L\"uscher term for the $\{6\}$ string approaches the value $-\pi/24$ of the fundamental
string for large separations of the color charges. However, at shorter distances, relevant
deviations from the asymptotic value are observed. This result could suggest that, for
large $N$, some intermediate regime may appear.

Concerning unstable strings, we have investigated the decay of confining strings in (2+1)-d
$SU(2)$ Yang-Mills theory. We have presented evidence for the decay of a string state
into another string state and for multiple decays resulting from the progressive partial
screening of color charges transforming under larger representations.

\acknowledgments
The results presented in this paper have been obtained in collaboration with F.~Gliozzi
and U.-J.~Wiese. I would like to thank them for many insightful discussions over the
years and for critical reading of this paper. I also thank A.~Amato for a useful remark.

\end{document}